\documentclass[journal]{IEEEtran}

\usepackage{graphicx}% Include figure files
\usepackage{dcolumn}% Align table columns on decimal point
\usepackage{bm}% bold math
\usepackage{units}
\usepackage[colorlinks=true,linkcolor=blue]{hyperref}%
\usepackage{siunitx}
\usepackage{cite}
\usepackage{amsmath}
\newcommand{\sub}[1]{\ensuremath{_{\mbox{\protect\scriptsize{#1}}}}}

\expandafter\ifx\csname package@font\endcsname\relax\else
 \expandafter\expandafter
 \expandafter\usepackage
 \expandafter\expandafter
 \expandafter{\csname package@font\endcsname}%
\fi
\def\mathclap#1{\text{\hbox to 0pt{\hss$\mathsurround=0pt#1$\hss}}} 
\newcommand{\msub}[1]{_{\text{#1}}}
\newcommand{\msup}[1]{^{\text{#1}}}

\newcommand{\avg}[1]{\left< #1 \right>} % for average

\DeclareFontFamily{U}{euc}{}% I chose euc because the chart is called Euler cursive 
\DeclareFontShape{U}{euc}{m}{n}{<-6>eurm5<6-8>eurm7<8->eurm10}{}% 
\DeclareSymbolFont{AMSc}{U}{euc}{m}{n} % I chose AMSc because AMSa and AMSb are defined in the amsfonts-package 
\DeclareMathSymbol{\umu}{\mathord}{AMSc}{"16}

\begin{document}

\title{Experimental evidence for electric surface resistance in niobium}

\author{\IEEEauthorblockN{T. Junginger,} \IEEEauthorblockA{Helmholtz-Zentrum Berlin f{\"u}r Materialien und Energie, Germany \\
Email: Tobias.Junginger@helmholtz-berlin.de \\}
 %\affiliation{CERN, Geneva, Switzerland}
 %\affiliation{MPIK Heidelberg, Germany}
%\and
\IEEEauthorblockN{S. Aull,} \IEEEauthorblockA{CERN, Geneva, Switzerland \\}
%\and
\IEEEauthorblockN{W. Weingarten,} \IEEEauthorblockA{retired from CERN, Geneva, Switzerland \\}%
 %\affiliation{CERN, Geneva, Switzerland}
%
%\and
\IEEEauthorblockN{C. P. Welsch,} \IEEEauthorblockA{Cockcroft Institute, Warrington and University of Liverpool, United Kingdom \\}}
%\affiliation{Cockcroft Institute, Warrington and University of Liverpool, United Kingdom}%\\This line break forced% with \\
%%}%

\maketitle

\begin{abstract}
Identifying the loss mechanisms of niobium cavities enables an accurate determination of applications for future accelerator projects and points to research topics required to mitigate current limitations. For several cavities an increasing surface resistance above a threshold field, saturating at higher field has been observed. Measurements on samples give evidence that this effect is caused by the surface electric field. The measured temperature and frequency dependence is consistent with a model that accounts for these losses by interface tunnel exchange between localized states in dielectric oxides and the adjacent superconductor. 
\end{abstract}


% \fi
\IEEEpeerreviewmaketitle

%\pacs{74.25.nn, 74.78.-w, 74.81.Bd}
\section{Introduction}
Superconducting cavities made of niobium are nowadays routinely reaching surface resistances $R\msub{S}$ as low as a few n$\Omega$ at surface magnetic fields above \unit[100]{mT} corresponding to peak electric fields of over 50 MV/m, some performing close to the theoretical limit of the material \cite{Padamsee:1180071}. Nevertheless many open questions concerning the field dependence of $R\msub{S}$ exist. Especially in the medium field region between a few and about \unit[100]{mT} differently prepared cavities show different field dependencies. Some cavities exhibit an increasing, some a decreasing surface resistance. Especially cavities prepared by coating a micrometer thick niobium film on a copper substrate exhibit a strong increase of $R\msub{S}$ with field. This paper focuses on cavities which show an increasing $R\msub{S}$ with field. 
%
%For some of these cavities this increase can be fitted with a polynomial of second order. These losses are described by models based on the surface magnetic field $\vec{B}$. Magnetic flux entry is thought to give rise to the quadratic term, which is dependent on temperature \cite{Wolfgangphysrev}, while the linear term is correlated to hysteresis losses and independent of temperature \cite{ciovatiHalbritter}. Adding an additional term to account for a widely observed decrease of $R\msub{S}$ at fields below a few mT \cite{Wolfgangphysrev} the total surface resistance for a cavity measured at a fixed temperature can be written as:
%\begin{equation}
%R\msub{S}=R_{0}+R_1\left(\frac{B}{B\msub{{c}}}\right)+R_2\left(\frac{B}{B\msub{c}}\right)^2+R_3\left(\frac{1}{B}\right)^2,
%\label{eq:polynomial}
%\end{equation}
%where the critical thermodynamic field $B\msub{c}$ is \unit[200]{mT} for niobium \cite{Leupold}. Even if these losses constitute the major contribution for most cavities, it is important to identify other loss types in order to disentangle them correctly. 
%
\section{Electrical losses from interface tunnel exchange}
Several authors have pointed out that there are several loss mechanisms involved which need to be addressed individually to obtain a better understanding and possibly mitigate their impact for specific cavities. Here we present a loss mechanism of electrical origin, already observed, though not further quantified, in prototypes of superconducting bulk niobium cavities for the Large Electron Positron Collider at CERN \cite{Bernard1981}. These losses yield an $R\msub{S}$ increasing above a threshold field saturating at higher field \cite{Jungingerphd} and can be explained by the interface tunnel exchange model (ITE) \cite{HR_electric_surface_impedance}. 
\begin{figure}[tbp]
   \centering
   \includegraphics[width=\columnwidth]{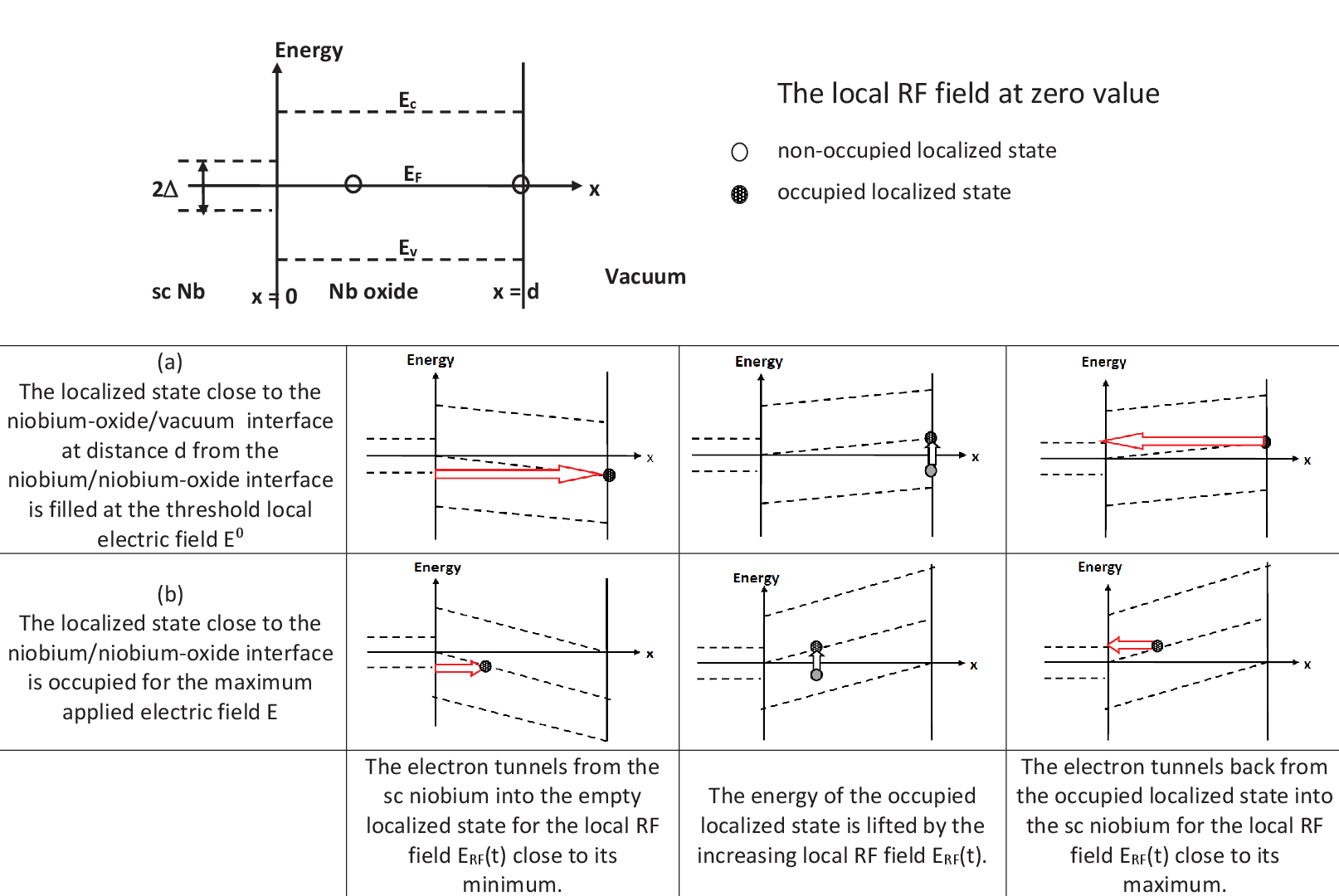}
   \caption{Current flow (red arrows) and voltage gain (black arrows) for (a) a localized state on top of the niobium-oxide layer  and (b) close to the niobium/niobium-oxide interface  (x - distance from niobium/niobium-oxide interface, $\Delta$ - energy gap of the sc niobium, $E\msub{F}$ - Fermi level,  , $E\msub{V}$ - Valence band of the niobium oxide, $E\msub{C}$ - conduction band of niobium oxide, d - thickness of niobium oxide layer). The RF field E lifts the occupied localized states above $E\msub{F}$+$\Delta$, from where they tunnel back into the sc niobium and dissipate the energy gained.}
   \label{figure:ITEWW}
\end{figure}
ITE assumes that electrons are exchanged between states in the superconducting Nb and localized states in adjacent dielectric oxides (Nb$_2$O$_5$ and/or NbO$_2$). 
%This exchange is caused by the surface electric field $\vec{E}$ penetrating only the oxide and not the superconductor, allowing for an exchange of electrons (i.e. a current) between the two materials within one RF period. Fig. \ref{figure:ITEWW} depicts the variation of the density of occupied electron states at zero field and after an exposure to an electric field. The time after exposure is considered here long as compared to the relaxation time. Hence the occupation of states is in thermal equilibrium. However, for sufficiently shorter times as in the case of a time-varying RF field the electron current density $\vec{j}$ is not in quadrature with the electric field, which gives rise to RF losses $\vec{j}\cdot \vec{E}$, and hence to an electric surface resistance $R\msub{S}\msup{E}$. 
This exchange is caused by the surface electric field $E$ penetrating only the oxide and not the superconductor, allowing for an exchange of electrons with gained energy (i.e. a current) by quantum mechanical tunneling between the niobium and its oxide (Fig. 1). Here we follow Halbritter \cite{HR_electric_surface_impedance} and take it as given that (i) the RF period is much shorter compared to the relaxation time of the occupied states in the oxide, (ii) the tunneling process is instantaneous, and (iii) the oxide thickness is much shorter than the range of the tunneling electrons. When the RF electric field is negative, the empty localized states are filled with electrons by quantum mechanical tunneling. The rising RF electric field lifts their potential energy until it exceeds $E\msub{F}+\Delta$ for a positive electric surface field. This threshold energy is reached first for the localized states far outside on top of the niobium-oxide, for which the voltage gain is maximum. The corresponding surface electric field is the threshold field $E\msub{0}$. From that moment on the electrons are tunneling back into the sc niobium, upon which they dissipate the energy gained. As the electrons relax and dissipate once per RF period $R\msub{S}\msup{E}$ depends linearly on the RF frequency $f$. Within the superconducting energy gap 2$\Delta$ there are no electronic states available for a current to flow. Therefore there exists a threshold electric field $E^0$, depending on the thickness of the oxide $d$ and on $\Delta$, below which there is no current and hence no RF loss. In a quantitative analysis, Halbritter calculated the surface resistance for ITE losses as \cite{HR_electric_surface_impedance}:
\begin{equation}
R\msup{E}\msub{S}=R\msup{E}\msub{S,sat}\left[\textrm{e}^{-b/E}-\textrm{e}^{-b/E^0}\right],\quad E\geq E^0.
	\label{eq:ITE}
\end{equation}
The parameters $R\msup{E}\msub{S,sat}$, $b$ and $E^0$ are defined by
\begin{equation}
b=\frac{2\kappa\Delta\varepsilon_r}{\beta^*e},\quad R\msup{E}\msub{S,sat}=\frac{2\pi f\umu_0}{(2\kappa)^2y}/f[\mathrm{GHz}],\quad E^0=\frac{\Delta\varepsilon_r}{ed\beta^*}
%\notag
\end{equation}
with
\begin{equation}
\kappa=\sqrt{2m\left(E\msub{c}-E\msub{F}\right)}/\hbar,\quad y^{-1}=\frac{\avg{xn\msub{T}}e^2}{\varepsilon\msub{0}\varepsilon\msub{r}}.
%\notag
\end{equation}
Here $R\msup{E}\msub{S,sat}$ is normalized to \unit[1]{GHz} to facilitate comparison of data sets obtained at different frequencies. The meaning of the physical parameters is the following: $E\msub{c}$ and $E\msub{F}$ are the energies of the conduction band and the Fermi energy, respectively; $\avg{xn\msub{T}}$ is the averaged product of the density of trapped electron states $n\msub{T}$ and the distance of the localized states from the niobium/niobium-oxide interface $x$, $d$ is the thickness of the oxide; $\varepsilon\msub{r}$ is the relative dielectric constant; $\beta^*$ is the geometric field enhancement factor of the metal due to surface roughness; $m$, $e$, $\varepsilon_0$, $\umu_0$ and $\hbar$ are the usual physical constants, such as the electron mass and electric charge, vacuum permittivity, vacuum permeability and Planck constant, in this order. The energy change in the oxide $Ex\beta^0$/$\varepsilon\msub{r}$ is in the order of a few meV enough to overcome the energy barrier.

\section{Experimental results}
\subsection{Tesla type bulk niobium cavity}
\begin{figure}[tbp]
   \centering
  \includegraphics[width=0.9\columnwidth]{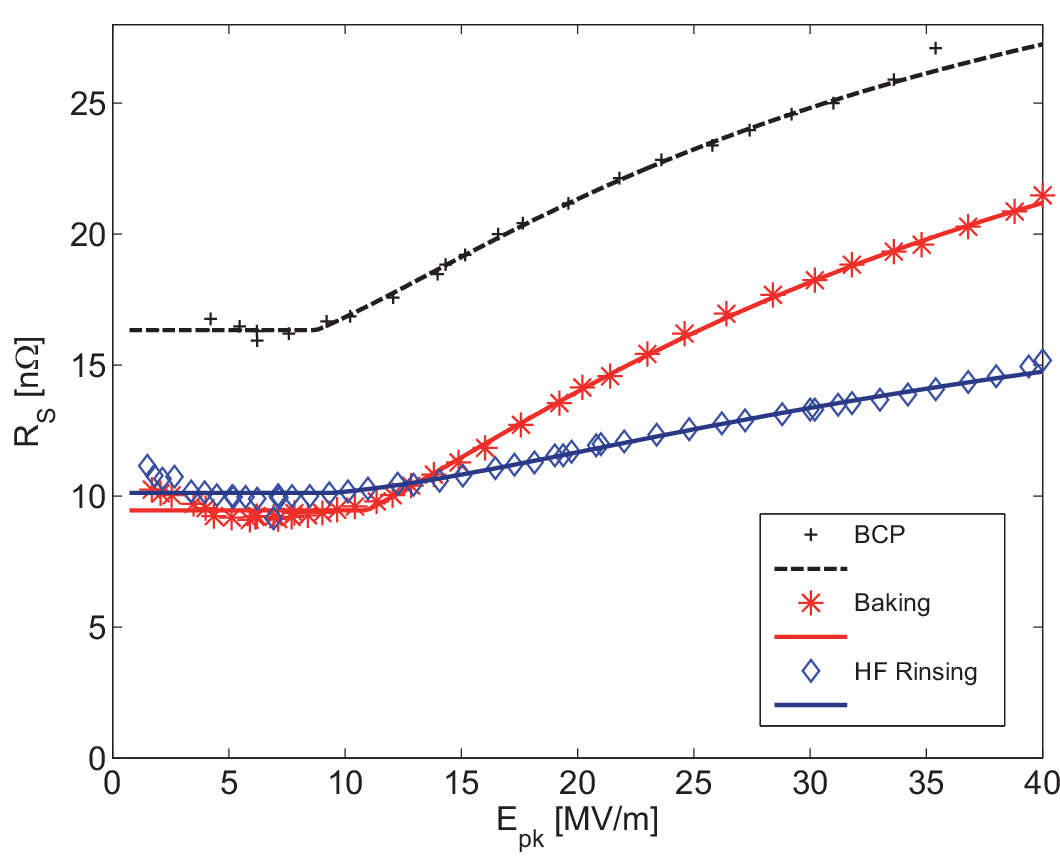}
   \caption{Surface resistance of an elliptical \unit[1300]{MHz} bulk niobium cavity at \unit[2]{K}. The lines show fits to the ITE model (Eq. \ref{eq:ITE2}. Data taken by Romanenko et al. \cite{Romanenko2013}} %The line shows the prediction from ITE model.}
   \label{figure:Alex}
\end{figure}

Figure \ref{figure:Alex} shows $R\msub{S}$ as a function of the peak electric field $E\msub{pk}$ at \unit[2]{K} of a \unit[1.3]{GHz} elliptical TESLA shaped \cite{PhysRevSTAB.3.092001} cavity. It was manufactured of fine grain bulk niobium (grain size of about \unit[50]{$\umu$m}). The first measurement was performed after chemical polishing (BCP). Afterwords the cavity was in situ baked at $120\,^{\circ}{\rm C}$. Then several  hydrofluoric (HF) rinsings were done to remove about \unit[10]{nm} of the outer surface layer \cite{Romanenko2013}. The dashed lines show fits to the ITE model with one additional parameter, an additional resistance $R\msub{0}$ which is assumed to be independent of field. This parameter accounts for losses from thermally activated particles and residual losses from other not further specified origin as well. For the analysis it is assumed that the field dependence is mainly due to ITE and therefore $R_0$ is taken as constant. The total $R\msub{S}$ is thus assumed to be
\begin{equation}
R\msub{S}=R\msup{E}\msub{S}(E)+R_{0}.
\label{eq:ITE2}
\end{equation}  
%where $R\msup{E}\msub{S}(E)$ is calculated according to Eq.\,\ref{eq:ITE}. For comparison the data has also been fitted to Eq. \ref{eq:polynomial}. 
%Note that the fit to Eq. \ref{eq:polynomial} systematically overestimates the data in the low and high field region, while it systematically underestimates the data in the medium field region for the data obtained after baking (red curve). The measurement is not well represented by the fit even if a relative high coefficient of determination $R^2$ is obtained. 
%The ITE model however can explain the dependence of $R\msub{S}$ on $E\msub{pk}$ better and significantly higher $R^2$ values are found for these fits obtained for the two data sets before HF rinsing, see Tab. \ref{tab:Fit_Parameters_Alex}. 

The phenomenological fit parameters (also found in Tab. \ref{tab:Fit_Parameters_Alex}) correspond to physical meaningful parameters (compare with \cite{HR87} and quotations therein and \cite{Delheusy,Kim}) as $\beta^*$=1, $\varepsilon_r$=10, $E\msub{c}-E\msub{F}$=\unit[0.05]{eV}, $\Delta$=\unit[1.18]{meV}, $d$=\unit[1.65]{nm} and $\avg{xn\msub{T}}$=\unit[6.7$\cdot10^{15}$]{1/(eVm$^2$)}, before and  $\beta^*$=1, $\varepsilon_r$=10, $E\msub{c}-E\msub{F}$=\unit[0.05]{eV}, $\Delta$=\unit[1.33]{meV}, $d$=\unit[1.27]{nm} and $\avg{xn\msub{T}}$=\unit[8.1$\cdot10^{15}$]{1/(eVm$^2$)} after baking. The values of $E\msub{c}-E\msub{F}$ are inconsistent with the band gap of Nb$_2$O$_5$ (3.4-5.3 eV \cite{Schultze}) but fit neatly the value of NbO$_2$ (0.1 eV \cite{Eyert}). Hence the localized states participating in the exchange could be found in the NbO$_2$ for which the value of $\varepsilon_r$=10 is consistent with \cite{Zhao}. Recent results show that after mild baking the total thickness of the oxide layer is reduced, but the thickness of the NbO$_2$ layer enhanced \cite{Delheusy}, which is consistent with an enhanced $R\msub{S,sat}\msup{E}$ and corresponding $\avg{xn\msub{T}}$. Another explanation for such a small $E\msub{c}-E\msub{F}$ is that the localized states are found at crystallographic shear planes in the Nb$_2$O$_5$. These are created by the NbO$_6$ building blocks sharing sides instead of indices \cite{HR87}. 
%
%\sout{Hence the localized states participating in the exchange are found the NbO$_2$ for which the value of $\varepsilon_r$=10 is consistent with}\cite{Zhao}.
%
\begin{table}[b]
   \centering
  \caption{Parameters derived for a least squares fit to Eq. \ref{eq:ITE2} of a bulk niobium cavity (cf. Fig. \ref{figure:Alex}).}
   \begin{tabular}{l|ccc}
        &BCP & $120\,^{\circ}{\rm C}$ baking  & HF rinsing  \\        
       \hline
         $R\msub{S,sat}\msup{E}$ in n$\Omega$ & 18.4$\pm$0.8 & 22.3$\pm$0.9 & 14.0$\pm$1.4 \\
				$E^0$ in MV/m & 7.1$\pm$1.2 & 10.5$\pm$0.5 & 4$\pm$250 \footnote{The threshold effect disappeared for this data set. The ITE model is not applicable here}   \\
       $b$ in MV/m & 26.9$\pm$1.2 & 30$\pm$3 & 44$\pm$4 \\
			$R_0$ in n$\Omega$ & 16.6$\pm$0.4  & 9.24$\pm$0.09& 11.0$\pm$0.2 \\
			%$R_3$ in n$\Omega$(mT)$^2$ &100$\pm$50 & 15$\pm$3& 3$\pm$3\\
			$R^2$ for Eq. \ref{eq:ITE2} & 0.9987& 0.9988 & 0.9885 \\
			%\hline
			%$R_0$ in n$\Omega$ &12.8$\pm$0.5 &6.9$\pm$0.8 & 10.4$\pm$0.2\\
			%$R_1$ in n$\Omega$ &38$\pm$2 &27$\pm$8 & 6.5$\pm$1.9\\
			%$R_2$ in n$\Omega$ & 0& 20$\pm$6&  19$\pm$4\\
			%$R_3$ in n$\Omega$(mT)$^2$ &170$\pm$80 &48$\pm$16 & 9$\pm$3\\
			%$R^2$ for Eq. \ref{eq:polynomial} & 0.9913& 0.9821&0.9926 \\

   \end{tabular}
   \label{tab:Fit_Parameters_Alex}
\end{table}
After HF rinsing the threshold seems to disappear. NbO$_2$ is well-known for its non-solubility in water and HF. This suggests that the states are rather located in the crystallographic shear planes of Nb$_2$O$_5$ and not in NbO$_2$.
%and the polynomial fit yields a better representation of the data.
%\sout{ITE losses require localized states inside a sufficiently thick dielectric (in this case NbO$_2$). Vanishing ITE after HF rinsing might be explained by a regrowth of a thinner fresh oxide layer with reduced NbO$_2$ content and localized states.}

In summary the analysis of the TESLA shaped cavity shows that there is a threshold effect as predicted by the ITE model in the surface resistance. The contribution of ITE to the field dependent surface resistance is enhanced by mild baking and mitigated by HF rinsing. This can be correlated to the oxide layer being altered by these processes. Inspecting Fig. \ref{figure:Alex} one might argue that the threshold field of the surface resistance $R\msub{S}$ at \unit[7-12]{MV/m} could be result of two opposing effects. For the first one $R\msub{S}$ may drop with the field, as observed by \cite{grassellino2013nitrogen} and for the second one $R\msub{S}$ may increase with field. We cannot refute this argument entirely. However, by virtue of the excellent fits of the ITE model to the data at higher fields its validity for the data before HF rinsing is suggested. 

\subsection{Niobium on Copper quarter wave cavity}  
\begin{figure}[tbp]
   \centering
  \includegraphics[width=0.9\columnwidth]{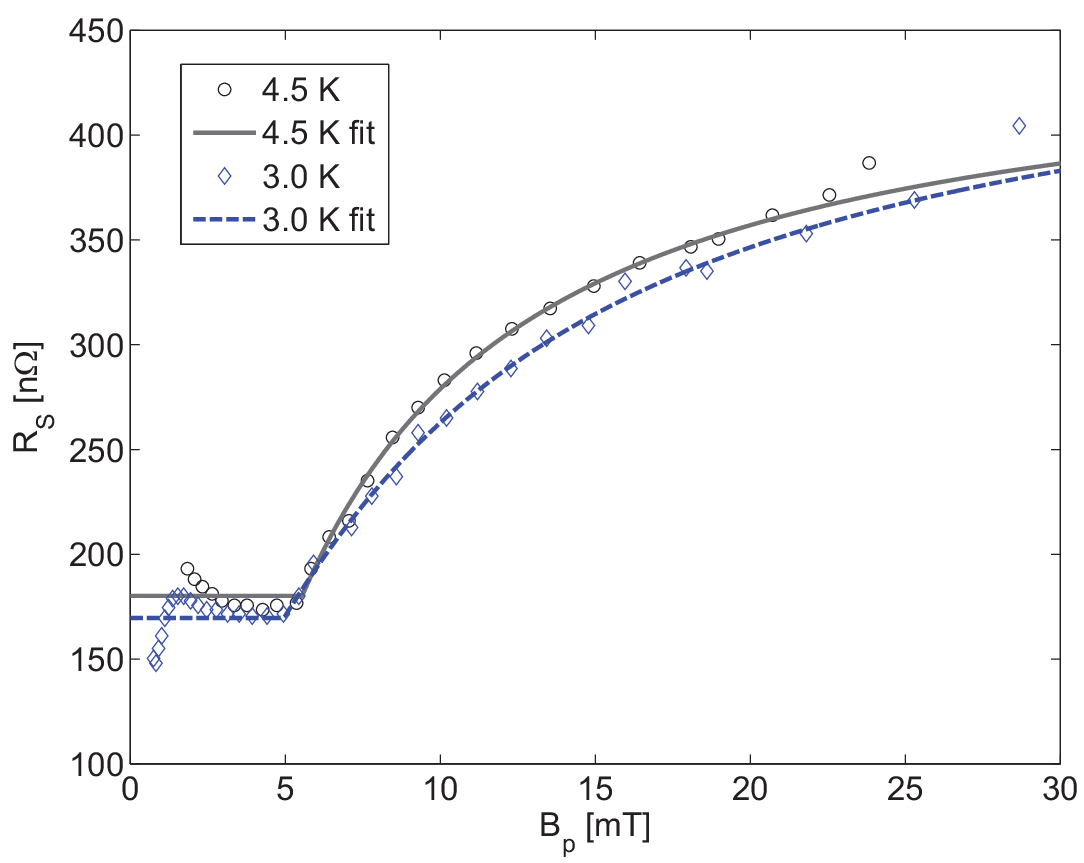}
  \caption{Surface resistance of a HIE-Isolde quarter-wave cavity at 3 and \unit[4.5]{K}. The uncertainty is about \unit[5]{\%} for each data point. Measurement were performed by M. Therasse and I. Mondino (CERN).}
   \label{figure:Isolde}
\end{figure}

The High Intensity and Energy On-Line Isotope Mass Separator (HIE-ISOLDE) is a facility currently under construction at CERN. It comprises a superconducting linac of quarter wave cavities produced by niobium on copper sputtering technology. Relevant cavity parameters for the analysis presented here are the magnetic and the electric geometry factor $G$=\unit[30.34]{n$\Omega$} and $G\msub{E}$=\unit[29.16]{n$\Omega$} respectively, the ratio between peak electric and magnetic field $E\msub{pk}/B\msub{pk}$=\unit[0.56]{MV/m} and the resonance frequency of \unit[101.28]{MHz}. In an early stage of the project a cavity has been produced which yielded a surface resistance exceeding the design value by almost a factor of ten, see Fig. \ref{figure:Isolde}. The surface resistance was measured at 3 and \unit[4.5]{K} to separate residual and BCS losses. Both curves show a threshold effect suggesting the ITE model is applicable here. The data for each temperate has been individually fitted to Eq. \ref{eq:ITE2}. Within the standard fit errors (95 \% confidence bounds), see Tab \ref{Tab:ITE_Isolde}, $R_0$ is equal for both temperatures showing that the losses are dominated by the residual resistance. The other phenomenological fit parameters $R\msub{S,sat}\msup{E}$, $b$ and $E^0$ can be correlated to a set of physical parameters with meaningful values as $\beta^*$=1, $\varepsilon_r$=10, $E\msub{c}-E\msub{F}$=\unit[0.01]{eV}, $\Delta$=\unit[1.04]{meV}, $d$=\unit[1.7]{nm} and $\avg{xn\msub{T}}$=\unit[3.6$\cdot10^{17}$]{1/(eVm$^2$)}. A critical assessment of these numbers lies however beyond the scope of this paper. The major difference compared to the TESLA type cavity is the larger $R\msub{S,sat}\msup{E}$ which can be explained by a larger value of trapped electron states $n\msub{T}$.   

\begin{table}[b]
   \centering
 \caption{Fit parameters interface tunnel exchange model to HIE-ISOLDE cavity data}
    \begin{tabular}{c|cccccc}

 $T$ in K& $R_0$[n$\Omega$] & $R\msub{S,sat}\msup{E}$[n$\Omega$] & $E^0$[MV/m]& $b$[MV/m] & $R^2$ \\ 
\hline 
3 & 193$\pm$3 &  6500$\pm$2200 & 5.4$\pm$0.2 & 3.5$\pm$1.6 & 0.9985\\ 

4.5 &188$\pm$4 & 4100$\pm$200 & 5.0$\pm$0.4 & 9$\pm$3 & 0.9973\\ 
  \end{tabular}
 \label{Tab:ITE_Isolde}
\end{table} 

In summary the analysis of the HIE-Isolde cavity confirms that there is a threshold effect as predicted by the ITE model in the surface resistance. The threshold field is independent of temperature as predicted by the model.

%
%Crack corrosion at grain boundaries can yield spots for tunneling, see Fig\,\ref{fig:ITE} (left). Losses due to interface tunnel exchange might therefore be more pronounced in sputtered cavities, since they have usually smaller grain sizes than bulk niobium cavities.
%
%
%The Quadrupole Resonator is a four-wire transmission line half-wave resonator using a TE$_{21}$-like mode. It was originally designed to measure the surface resistance of niobium film samples at \unit[400]{MHz}, the technology and RF frequency chosen for the LHC at CERN. The device was used to measure the surface resistance at \unit[400]{MHz} of bulk niobium and niobium on copper samples \cite{Brigant:360751,Chiaveri:360555,Mahner:611593} in the past. In the framework of this thesis it has been refurbished in order to extend the measurement range to 800 and \unit[1200]{MHz} and to probe the critical field of the samples \cite{Junginger:SRF09,Junginger:SRF11,Junginger:IPAC}.
%

\subsection{Quadrupole Resonator measurements}  
In order to test whether the losses accounted for by ITE scale linearly with frequency as predicted by the model a cavity test is not suitable. The Quadrupole Resonator \cite{Mahner:611593} is a unique device enabling to test $R\msub{S}$ of superconducting samples over a wide parameter range. It features two excitable modes at 400 and \unit[800]{MHz} with identical magnetic field configuration on the sample surface. The ratio between electric and magnetic field for these two modes is proportional to $f$. For example for a peak magnetic field  $B\msub{p}$=\unit[10]{mT}, the peak electric fields are $E\msub{p}$=\unit[0.52 and \,1.04]{MV/m} for \unit[400 and 800]{MHz}, respectively \cite{Junginger:Revscientinstr12}. This feature allows for a separation of magnetic and electrical losses from measurement data by comparison with theoretical models as will be explained and carried out in the following. When analyzing surface resistance data obtained with the Quadrupole Resonator several things have to be taken into account to compare the results to cavity data. One advantage is that the heat flow on the sample surface is completely lateral and therefore there is no thermal feedback effect \cite{Junginger:Revscientinstr12}. The field configuration of accelerating cavities is in general such that the magnetic and the electric geometry factor are almost equal. For the sample surface area of the Quadrupole Resonator this is not the case. By the calorimetric (RF-DC compensation) technique explained in detail in \cite{Junginger:Revscientinstr12} one obtains the power dissipated under RF $P\msub{RF}$ directly. Using the field distribution on the sample and in the entire resonator volume one can derive the magnetic surface resistance $R\msub{S}$ of the sample using     

\begin{equation}
R\msub{S}=\frac{2\mu_0^2P\msub{RF}}{\int\limits_{{\mathrm{Sample}}} |\vec{B}|^2\mathrm{d}S}.
\label{eq:Rmag}
\end{equation}

This formula assumes that all losses are caused by the surface magnetic field. If however one would assume that all losses are caused by the surface electric field, 

\begin{equation}
R\msub{S}\msup{E}=\frac{2\mu_0P\msub{RF}}{\varepsilon_0\int\limits_{{\mathrm{Sample}}}|\vec{E}|^2\mathrm{d}S}.
\label{eq:Rel}
\end{equation}
has to be used to calculate the electric surface resistance $R\msub{S}\msup{E}$. For most accelerating cavities $R\msub{S}\approx R\msub{S}\msup{E}$ holds. Note that this does not mean that the losses from electric and magnetic origin are equal, but that the magnetic and the electric geometry factors are equal. For the Quadrupole Resonator with its narrow gap structure $R\msub{S}\approx R\msub{S}\msup{E}$ does not hold. For example a power $P\msub{RF}$ dissipated on the sample at \unit[400]{MHz} corresponding to an $R\msub{S}$ of \unit[1]{n$\Omega$} corresponds to an $R\msub{S}\msup{E}$ of \unit[53.5]{n$\Omega$}. An interesting feature of the Quadrupole Resonator is that the ratio between electric and magnetic field is frequency dependent and scales like    

\begin{equation}
{\int\limits_{{\mathrm{Sample}}}|\vec{E}|^2\mathrm{d}S}/{\int\limits_{{\mathrm{Sample}}}|\vec{B}|^2\mathrm{d}S} \propto f^2.
\label{eq:f2}
\end{equation}
This means that a power $P\msub{RF}$ dissipated at \unit[800]{MHz} corresponding to an $R\msub{S}$ of \unit[1]{n$\Omega$} corresponds to an $R\msub{S}\msup{E}$ of \unit[13.4]{n$\Omega$}. For details see \cite{Jungingerphd}.

%However it is not possible to alter the electric and magnetic field independently. First one has to assume that the losses are completely caused by only one of the fields.

\begin{figure}[t]
   \centering
   \includegraphics[width=0.95\columnwidth]{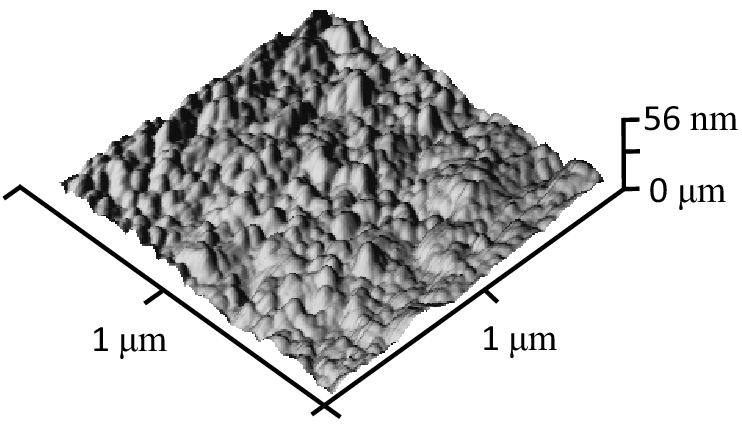}
   \caption{Surface profile from the niobium film sample obtained from AFM. The lateral resolution of the image is \unit[4]{nm}.}
	\label{fig:AFM}
\end{figure}

To test the properties of ITE a sample is required for which these temperature independent losses remain dominant up to relatively large temperatures. This condition was obtained for a micrometer thin niobium film sample sputtered on a copper substrate, which was kept under normal air for 10 years. Using XPS the thickness of its surface layer was found be significantly larger as a reference bulk niobium sample prepared by BCP \cite{Jungingerphd}. The thin film has a grain size of a few nm as measured by atomic force microscopy see Fig \ref{fig:AFM}. This is several orders of magnitude smaller than typical values of fine grain bulk niobium surfaces prepared for accelerating cavities. Oxides formed between grain boundaries can significantly increase the interface area through which the tunneling process occurs. 

Using the Quadrupole Resonator the sample was measured at 400 and \unit[800]{MHz} over a temperature range between 2 and \unit[4.5]{K}. 
%Figure \ref{figure:RB_NbCU} shows $R\msub{S}$ vs. the peak magnetic field $B\msub{pk}$ on the sample. 
In general the Quadrupole Resonator enables also measurements at \unit[1200]{MHz}. However for the sample investigated here thermal runaway was observed for fields above about \unit[10]{mT}. That is why no data measured at \unit[1200]{MHz} is included in the analysis. The complete data set consists of 183 values $R\msub{S}(f,T,B,E)$. It has been collectively fitted to a single set of parameters. In order to do this Eq.\ref{eq:ITE3} had to be extended to account for the BCS losses for each temperature and frequency. The total surface resistance is now considered to be
\begin{equation}
R\msub{S}=R\msup{E}\msub{S}(E)+R_{0}(f)+R\msub{BCS}(f,T).
\label{eq:ITE3}
\end{equation}  
Unlike for the analysis above here $R_{0}$ only accounts for the residual losses from sources other than ITE. BCS losses from thermally activated particles are accounted for by $R\msub{BCS}(f,T)$. The field dependence of both terms are considered small compared to the ITE losses. For the BCS surface resistance 
\begin{equation}
R\msub{BCS}=\mu_0^2 \omega^2 \sigma_0 RRR\cdot \lambda(T,l)^3 \frac{\Delta}{k_B T} \ln{\left(\frac{\Delta}{\hbar\omega}\right)} \frac{e^{-\Delta/k_BT}}{T}
     \label{eq:BCS}
\end{equation} 
is used, which is a good approximation in the dirty limit. Low field surface resistance and penetration depth change measurements have shown that this sample has a low RRR and therefore the dirty limit approximation is reasonable \cite{Jungingerphd}. No parameter of $R\msub{BCS}$ is varied to minimize $\chi^2$. All are obtained from different measurements. While RRR and $\Delta$ are taken from low field surface resistance measurements, $\lambda$ is taken from penetration depth change measurements. For details refer to \cite{Jungingerphd}.  

The complete data set consisting of 183 values $R\msub{S}(f,T,E)$ has been collectively fitted to Eq. \ref{eq:ITE3} with five fit parameters. A $\chi^2$=167.9 was obtained for the fit parameter values of $R\msub{S,sat}\msup{E}$=\unit[17000$\pm$500]{n$\Omega$}, $b$=\unit[1.06$\pm$0.10]{MV/m} and $E^0$=\unit[0.610$\pm$0.015]{MV/m}, $R_0$=\unit[275$\pm7$]{n$\Omega$} at \unit[400]{MHz} and $R_0$=\unit[500$\pm11$]{n$\Omega$} at \unit[800]{MHz}. The value of $\chi^2$ is slightly lower than the number of data points indicating that the experimental uncertainty was a bit overestimated.

\begin{figure}[t]
   \centering
   \includegraphics[width=0.95\columnwidth]{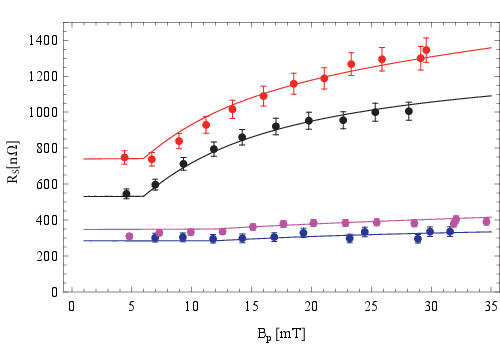}
   \caption{Surface resistance $R\msub{S}$ of a niobium film sample tested with the Quadrupole Resonator at \unit[400]{MHz} (\unit[2.5]{K} (blue), \unit[4]{K} (magenta)) and \unit[800]{MHz} (\unit[2.5]{K} (black), \unit[4]{K} (red). Each data point $R\msub{S}$ was obtained under the assumption that all losses are solely caused by the surface magnetic field. The lines show predictions from a collective least squares multiparameter fit to Eq. \ref{eq:ITE3}. The total data set comprises 183 values $R\msub{S}(f,T,B,E)$.}
   \label{figure:RB_NbCU}
\end{figure}

\begin{figure}[t]
   \centering
   \includegraphics[width=0.95\columnwidth]{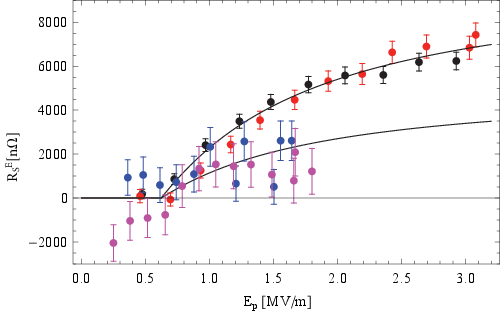}
   \caption{Electric surface resistance at \unit[400]{MHz} (\unit[2.5]{K} (blue), \unit[4]{K} (magenta)) and \unit[800]{MHz} (\unit[2.5]{K} (black), \unit[4]{K} (red)) of a niobium film sample. The same data is plotted as in Fig. \ref{figure:RB_NbCU}, here under the assumption that all field dependent losses are caused by the surface electric field. $R\msub{BCS}$ and $R_0$ have been subtracted from each data point and the fit curves.}
   \label{figure:RE_NbCU}
\end{figure}

For illustration $R\msub{S}$ is plotted as a function of $B$ for the temperatures of 2.5 and \unit[4]{K} for both frequencies in Fig. \ref{figure:RB_NbCU}. This corresponds to about one fifth of the complete data set. Note that to calculate $R\msub{S}$ for each individual data point Eq. \ref{eq:Rmag} has been used. To fit the complete data collectively to a single set of parameters the frequency dependence \ref{eq:f2} has to be taken into account. Subtracting the fitted values of $R_{0}(f)$ and the calculated $R\msub{BCS}(f,T)$ individually from each data point $R\msup{E}\msub{S}(E)$ can be calculated from Eq. \ref{eq:Rel}. This data is plotted in Fig. \ref{figure:RE_NbCU}.     


For this sample $R\msub{S,Sat}\msup{E}$ is three orders of magnitude larger than for the bulk niobium TESLA type cavity, corresponding to a higher density of trapped states similar to what has been found for the HIE-Isolde quarter wave cavity also produced from sputtering niobium on a copper substrate. The onset field $E^0$ is one order of magnitude smaller for the sample, which might be correlated to the roughness of the sample in the nanometer scale, as measured by AFM, see Fig \ref{fig:AFM}. For further surface analytic measurements on this sample in comparison to bulk niobium surfaces refer to \cite{Jungingerphd}. Also here, the phenomenological fit parameters $R\msub{S,sat}\msup{E}$, $b$ and $E^0$ can be correlated to a set of physical parameters with meaningful values as $\beta^*$=12, $\varepsilon_r$=10, $E\msub{c}-E\msub{F}$=\unit[0.01]{eV}, $d$=\unit[2.0]{nm} and $\avg{xn\msub{T}}$=\unit[1.2$\cdot10^{18}$]{1/(eVm$^2$)}. A critical assessment of these numbers lies however beyond the scope of this paper.

In summary the analysis of the niobium on copper sample measured with the Quadrupole Resonator confirms the threshold effect observed for the cavity data shown above. It is also confirmed that this threshold does not depend on temperature. The strong frequency dependence of $R\msub{S}$ observed can be explained by electrical losses taking into account the Quadrupole Resonator's frequency dependent ratio between electric and surface magnetic fields. A collective fit to the ITE model of a large data set comprising two frequencies and several temperatures gives an excellent representation of the data suggesting the ITE model is indeed applicable to this data.   

\section{Discussion}
A variety of surface resistance data from different cavities and samples have been fitted to the ITE model. The Quadrupole Resonator measurements showed smaller threshold fields $E_0$ compared to the cavity data. This variation can for the most part be attributed quite naturally to a variation of the oxide thickness with the applied treatment. X-ray photoelectron spectroscopy (XPS) measurements on this sample have shown that the oxide layer of this sample is thick compared to a bulk niobium sample for which no ITE losses have been measured \cite{Jungingerphd}. Also in \cite{Jungingerphd} the data has been analyzed with different theoretical models. The huge difference in the field dependent $R\msub{S}$ for the different frequencies in combination with the frequency dependent ratio $B\msub{p}/E\msub{p}$ strongly suggests an electrical loss mechanism. 

\subsection{Comparison with the thermal contact resistance model}

The thermal contact resistance model was recently proposed to explain the field dependent losses in Nb/Cu cavities and samples \cite{Vaglio_TBR}.
The basic idea of the model is that the superconducting film is not perfectly in contact with the copper substrate but locally detached without peeling off. Under the influence of the RF field, these microscopic spots heat up due to the suppressed cooling until they go into a local thermal runaway and are driven into the normal conducting state. 
The stronger the RF field, the more micro-quenches occur and contribute to the overall surface resistance with their normal resistivity.
Each overall performance curve $\overline{R\sub{S}}$ can be described by the local surface resistance $R\sub{S}$ and a statistical distribution function of contact resistance $f(R\sub{B})$:

\begin{equation}\label{eq:inverseproblem}
\overline{R\sub{S} \left(T\sub{0},B\sub{p}\right) } = \int_0^\infty R\sub{S}\left(T\sub{0},B\sub{p},R\sub{B}\right)f(R\sub{B}) dR\sub{B}.
\end{equation}

$\overline{R\sub{S}}$ is derived from measurements via measuring the quality factor $Q$ for cavities (with $\overline{R\sub{S}} = G/Q$) or via a calorimetric method for Quadrupole Resonator samples.
A distribution function can then be derived by inverting Eq. \ref{eq:inverseproblem}.

\begin{figure}
\centering
\includegraphics[width=1\linewidth]{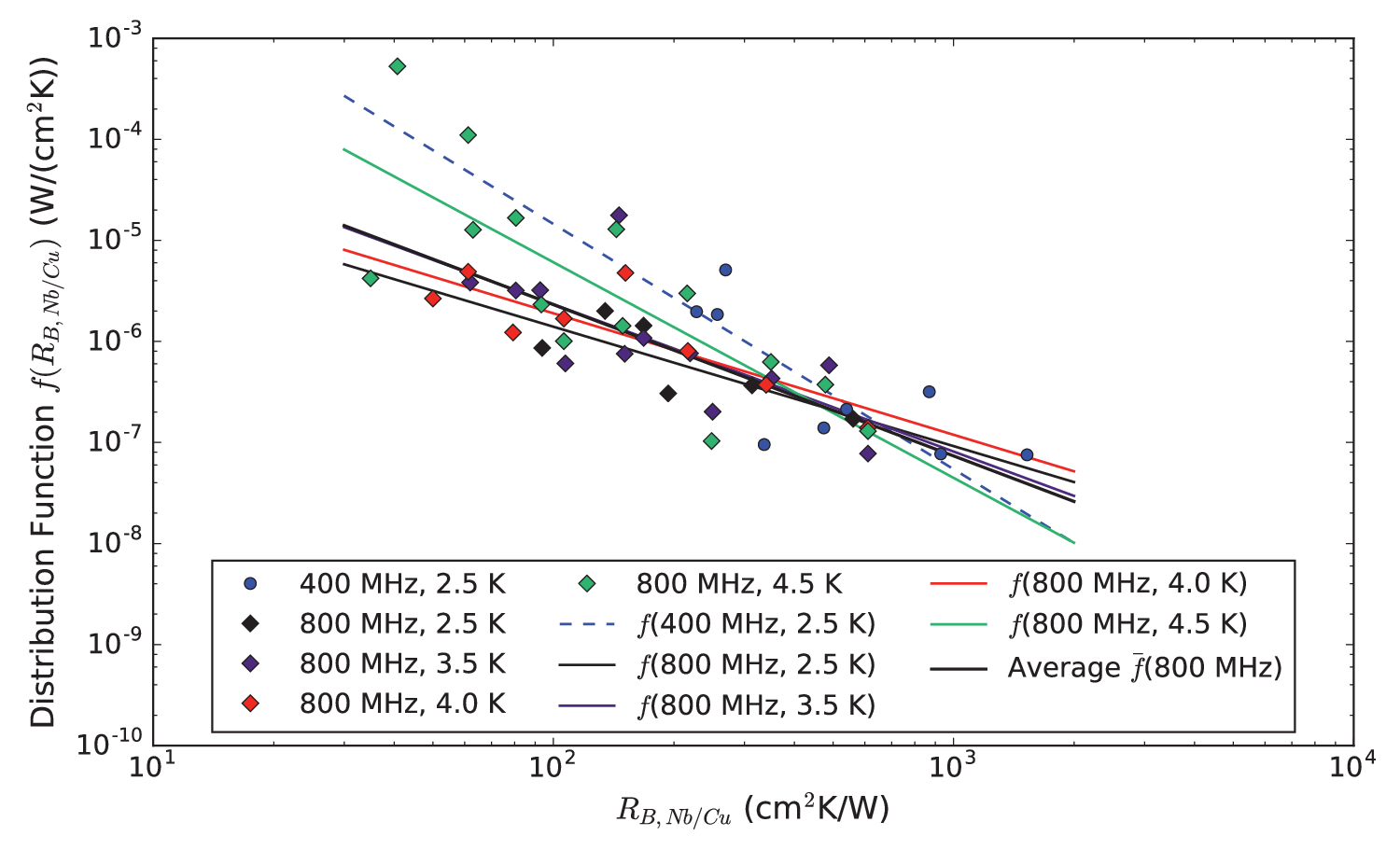}
\caption{Distribution function $f(R\sub{B, Nb/Cu})$ for the niobium film sample tested with the Quadrupole Resonator. The corresponding surface resistance is shown in Fig. \ref{figure:RB_NbCU}.}\label{fig:DistributionFunction}
\end{figure}

We process the data of the niobium film sample measured in the Quadrupole Resonator in the context of this model: 
Based on the $R\sub{S}(T)$ measurements the temperature increase $\Delta T$ for a given contact resistance is calculated 
via
\begin{equation}
\Delta T = R\sub{B} \frac{1}{2} R\sub{S} \left(\frac{B\sub{p}}{\mu\sub{0}}\right)^2
%\Delta T = R\sub{B} \frac{1}{2} R\sub{S} H\sub{p}^2
\end{equation}
and fed back into $R\sub{S}(T) \rightarrow R\sub{S}(T+\Delta T)$ for each frequency and bath temperature $T\sub{0}$ combination.
As a result the thermal runaway field $B\sub{quench}$ is derived as function of contact resistance values which can then be used to translate 
$R\sub{S}\left(B\sub{p},R\sub{B}\right)$ into $R\sub{S}\left(B\sub{p},B\sub{quench}\right)$. 
We follow \cite{Vaglio_TBR} where the integral in Eq. \ref{eq:inverseproblem} is approximated by appropriate
discretization procedures and use $R\sub{S}\left(B\sub{p},B\sub{quench}\right)$ to obtain a set of distribution function points which can then be fitted with an anlaytic expression. We process all available data sets, but discard sets with less than 6 points for the distribution function and sets where the standard error of the fit is bigger than \SI{60}{\%}. 
The remaining data sets along with their distribution function are shown in Fig. \ref{fig:DistributionFunction}.
The distribution functions of the \SI{800}{MHz} data agree moderately well with each other; however with large standard errors.
We calculate the weighted average of the distribution functions of the \SI{800}{MHz} data across all temperatures and find that the $f\left(\SI{800}{MHz},\SI{4.5}{K}\right)$ even lies outside the weighted standard deviation $\overline{f}$ .

The model also allows for an estimate of the corresponding surface fraction of detached film from the substrate via
\begin{equation}
A\sub{detached} = \int_{\text{min}(R\sub{B})}^\infty f\left(R\sub{B, Nb/Cu}\right)dR\sub{B, Nb/Cu}.
\end{equation}
For the \SI{800}{MHz} data we find on average
\begin{equation}
A\sub{800,detached} = \left(0.058 \pm 0.008 \right)\si{\%}
\end{equation}
which is the typical order of magnitude \cite{Vaglio_TBR}.

From the \SI{400}{MHz} data sets only the \SI{2.5}{K} set remained. 
The resulting distribution function is similar to the (\SI{800}{MHz}, \SI{4.5}{K}) data but significantly different from all other data sets.
Also the detached surface area of \SI{0.0025}{\%} is not in agreement with the \SI{800}{MHz} results.

It has to be noted that the thermal contact resistance model is not a fit and a solution can be found for any data set.
Hence, to conclude that the thermal contact resistance model is the dominant loss mechanism, the same distribution function ought to be derived for different temperatures and/or frequencies.
This has been consistently shown with previous analyses on standard performing HIE-ISOLDE cavity data at different temperatures \cite{Miyazaki_TBR} and on another Quadrupole Resonator Nb/Cu sample at different frequencies and temperatures \cite{Aull_TBR}.
Given the large errors and only moderate agreement of the presented data, we consider the analysis not supporting the thermal contact resistance model as the dominant loss mechanism.
Moreover, given the nature of the loss mechanism, we would not expect a saturation but rather a stronger surface resistance increase with higher RF fields as often observed as an exponential increase of $R\sub{S}$ with field in Nb/Cu cavities.
Overall, we conclude that the analysis of the data in the context of the thermal contact resistance model neither contradicts nor sufficiently competes with ITE being the dominant loss mechanism.

\section{Conclusion}
In conclusion, the dependency of the  surface resistance on the applied field strength strongly depends on the surface preparation. This indicates a variety of different dominant field dependent loss mechanisms. Some cavities exhibit an $R\msub{S}$ increasing above a threshold field saturating at higher field. In this paper it has been shown that measurements on a bulk niobium cavity, showing this behavior of $R\msub{S}$ on the surface electric field, can be well described by the ITE model. For a niobium on copper quarter wave cavity these losses were found to be stronger, which could be correlated to more trapped states participating in the exchange. To further test the predictions of the ITE model a niobium thin film sample was tested with the Quadrupole Resonator. These measurements showed field dependent losses independent on temperature, which scale linearly with frequency, if one assumes that they are caused by the surface electric field. These findings are consistent with the predictions of the ITE model and in contradiction with the thermal contact resistance model.

Our results allow to better understand the field dependent surface resistance of superconducting niobium. This can be used for the development of future accelerating cavities. In particular a possible explanation for the larger field dependent surface resistance found in some cavities produced of niobium films on copper substrates, a technology widely used for cavity operation at \unit[4.2]{K} \cite{Sergio2006}, is given by the ITE model.

\section{Acknowledgment}
The authors would like to thank J\"urgen Halbritter for fruitful discussions on the ITE loss mechanism, A. Romanenko, I. Mondino and M. Therasse for providing the cavity data. This work was supported by the German Doctoral Students program of the Federal Ministry of Education and Research (BMBF) and by a Marie Curie International Outgoing Fellowship within the EU Seventh Framework Programme for Research and Technological Development (2007-2013).
%The growth of a natural surface layer of several nanoohms is only responsible for rather small residual losses 

%The electrical field $E$ on the Quadrupole Resonator sample surface scales linearly with frequency for a given magnetic field $B$, as required by the law of induction as applied to the geometry in between the crooked ending of the rods and the sample. For a peak magnetic field  $B\msub{p}$=\unit[10]{mT}, the peak electric field is $E\msub{p}=$\unit[0.52(1.04)]{MV/m} for \unit[400(800)]{MHz}, respectively, as previously mentioned. That is why, the data at \unit[400]{MHz} was only measured for electric fields below \unit[2]{MV/m}, which corresponds to \unit[35.0]{mT}. Measurements at higher field level were prevented by thermal runaway. 
%
%The residuals seem to be evenly distributed. No apparent trend, that data is  systematically over- or underestimated for one frequency or a specific area of the surface resistance can be found in Fig. \ref{fig:Residualsniobiumfilm}. The residuals are systematically smaller at \unit[800]{MHz} than at \unit[400]{MHz} indicating that the model is better suited to describe the electric than the magnetic losses, which are only dominant at the higher frequency.  

%\bibliography{ApplPhysLett}
%\bibliographystyle{IEEEtran}  

\end{document}